# Schrödinger dynamics as a two-phase conserved flow: an alternative trajectory construction of quantum propagation

**Peter Holland**

Green Templeton College
University of Oxford
Woodstock Road
Oxford OX2 6HG
England

peter.holland@gtc.ox.ac.uk

12th December 2008

**Abstract**

It is shown that the Schrödinger equation can be cast in the form of two coupled real conservation equations, in Euclidean spacetime in the free case and in a five-dimensional Eisenhart geometry in the presence of an external potential. This implies a novel two-phase quantum hydrodynamic model whose Lagrangian picture provides an exact scheme to calculate the time-dependent wavefunction from a continuum of deterministic trajectories where two points are linked by at most two trajectories. Properties of the model are examined, including the appearance of 'entangled' trajectories in separable states. Wavefunction constructions employing alternative two-phase models are proposed.

PACS: 03.65.Ca , 03.65.Ta, 47.55.-t

## 1. Introduction

The hydrodynamic analogy has provided fruitful insights in several areas of quantum mechanics (e.g., [1]). One of the earliest such approaches was due to Madelung [2] who observed, in effect, that wave mechanics could be regarded as the Eulerian picture of a hydrodynamic model in which quantum effects are encoded in a generalization of Euler's force law. The success of the analogy derives from the fact that local conservation laws play a primary role in both disciplines (quantum and hydrodynamic). The analogy has influenced a significant strand of interpretational studies and its close connection with the de Broglie-Bohm theory was noted early on [3]. It has been appreciated only recently, however, that the association of the de Broglie-Bohm approach with fluid mechanics yields a significant computational benefit: (a) the de Broglie-Bohm trajectories[1] can be computed independently of the wavefunction (only the initial wavefunction is needed) and (b) they exhibit sufficient structure to provide a method to generate the time-dependence of the wavefunction. The trajectory viewpoint is just the Lagrangian picture [4] of this quantum-hydrodynamical model and it is the Lagrangian expression of conservation that provides the key in ascribing to the trajectory a constructive role in quantum propagation [5][2].

---

[1] The terms 'trajectory' and 'path' are used interchangeably.

Much valuable computational work has been done within this scheme, a well developed approach being the 'synthetic' method where the wave and trajectory equations are solved simultaneously [6]. It is natural to enquire whether there are yet more quantum trajectory constructions that exploit the association with continuum mechanics. This possibility is implicit in modifications of the de Broglie-Bohm model that utilize the freedom to define alternative laws for the paths which are compatible with the same density [7] but the question may be posed in other ways. For example, is it necessary in a hydrodynamic approach that pairs of spacetime points are linked by at most one deterministic path (a characteristic of the de Broglie-Bohm flow), or might quantum propagation be described using a fixed, finite number (greater than one) of deterministic connecting paths? In this contribution we shall show that there is indeed another option: the wavefunction may be built from a patchwork of trajectories in which spacetime points are linked by (at most) two paths.

The trajectory construction presented here, which exploits the continuity equation in a novel way, follows from a simple but generally unappreciated property of the Schrödinger equation,

$$i\hbar \frac{\partial \psi}{\partial t} = -\frac{\hbar^2}{2m} \nabla^2 \psi + V\psi, \qquad (1.1)$$

namely, that (1.1) implies conservation laws not only for quadratic and other multiplicative combinations of the wavefunction, but also for the (linear) wavefunction itself. This property is obvious in the free case and we show how to extend the result to the non-free case by going to five dimensions. We thereby represent quantum evolution through a pair of (real) Eulerian conservation equations that, in fluid parlance, defines a two-phase flow. The corresponding Lagrangian picture implies two sets of congruences from which the temporal evolution of the wavefunction may be derived. As we shall see, this quantum hydrodynamic model differs in key respects from the Madelung/de Broglie-Bohm model.

**2. Multiphase flow**

---

[2] Note that we are employing only the trajectories of the de Broglie-Bohm theory and not its theory of matter, in which one path is labelled preferentially and occupied by a material corpuscle. Here, *all* the paths that are potentially available to the de Broglie-Bohm corpuscle, only one of which is traversed according to that theory, are employed simultaneously in describing quantum propagation. And the notion of forces acting between the paths, fundamental to the constructive model, is not meaningful in the de Broglie-Bohm theory.

We shall need the following elementary facts about conserved flows. Consider a continuous dynamical system whose configuration is described by a set of $n$ Cartesian coordinates $x_i, i = 1,...,n,$ and the time $t$. In the Eulerian viewpoint, the system is characterized by an associated set of single-valued real fields $\rho_a(x,t)$ and $v_{ai}(x,t)$, $a = 1,...,N$. For each $a$, the function $\rho_a$ is a density (which may take negative values, such as electric charge) and $v_{ai}$ is a corresponding velocity. For $a > 1$ these functions describe a multiphase flow of $N$ interpenetrating fluids (for details see, e.g., [8-13]).

In the Lagrangian viewpoint, we describe the history of the $a$th phase by the paths $q_{ai}(q_{a0},t)$ of each fluid particle[3]. Here $q_{a0i}$ are the initial coordinates with which it is convenient to label the particles. The fundamental physical quantities in the Lagrangian theory of motion, which describe interparticle forces, are the deformation matrices $J_{aii'} = \partial q_{ai}/\partial q_{a0i'}$. By the inverse function theorem the trajectory equation is invertible locally: $q_{a0i} = q_{a0i}(q_a,t)$ if $\det J_a > 0$. In a multiphase flow each space point supports simultaneously a fluid particle of each phase $a$. Denoting the density in the reference state by $\rho_{a0}(q_{a0})$, a fundamental characteristic of the flow is that each phase $a$ obeys a local conservation law:

$$\frac{d}{dt}\left[\rho_a(q_{a0},t)d^n q_a(q_{a0},t)\right] = 0 \tag{2.1}$$

or

$$\rho_a(q_{a0},t)d^n q_a(q_{a0},t) = \rho_{a0}(q_{a0})d^n q_{a0}. \tag{2.2}$$

We may deduce from this that the corresponding Eulerian functions[4],

$$\rho_a(x,t) = \rho_a(q_{a0},t)\big|_{q_{a0}(x,t)}, \quad a = 1,...,N, \tag{2.3}$$

$$v_{ai}(x,t) = \frac{\partial q_{ai}(q_{a0},t)}{\partial t}\bigg|_{q_{a0}(x,t)}, \quad a = 1,...,N, \tag{2.4}$$

obey the equations[5]

$$\frac{\partial \rho_a(x)}{\partial t} + \sum_{i=1}^{n}\frac{\partial}{\partial x_i}\left[\rho_a(x)v_{ai}(x)\right] = 0. \tag{2.5}$$

---

[3] The term 'particle' has no ontological significance in the present context.
[4] This is poor notation since the Eulerian and Lagrangian density functions are in general different functions of $t$. We follow conventional practice and distinguish them by their arguments.
[5] Throughout this paper summation on repeated indices is indicated explicitly.

Eqs. (2.1) and (2.5) are equivalent statements of local conservation as portrayed in the Lagrangian and Eulerian pictures, respectively. Eqs. (2.2) and (2.3) therefore give the general solution of (2.5) in terms of the paths, a result due to Euler [8]. We shall exploit this connection in the following way. Conventional field theories describe physical processes in terms of functions of fixed space points, i.e., in a language that corresponds to the Eulerian picture in continuum mechanics. If we have a field theory involving an equation of the sort (2.5), we may seek to associate with it a fluid-dynamical model with respect to which this equation is an Eulerian-picture local conservation law for some quantity. Naturally, the appropriateness of this step needs to be checked in each case but it is noteworthy that many equations of physics have the conservation form [14] (although this fact has not been widely exploited). The key insight provided by the continuum-mechanical approach is that one may then attempt to invoke the corresponding Lagrangian picture according to which the time-dependent $a$th density $\rho_a(q_{a0},t)$ may be constructed from the $a$th set of trajectories using (2.2). Then, in a field theory that admits a suitable equation (2.5), the Eulerian density is given by (2.3), i.e.,

$$\rho_a(x,t) = \int \delta(x - q_a(q_{a0},t)) \rho_{a0}(q_{a0}) d^n q_{a0} \tag{2.6}$$

or

$$\rho_a(x,t) = J_a^{-1}(q_{a0},t) \rho_{a0}(q_{a0}) \big|_{q_{a0}(x,t)}, \quad a = 1,...,N, \tag{2.7}$$

where

$$J_a(q_{a0},t) = \det \frac{\partial q_a}{\partial q_{a0}}. \tag{2.8}$$

This determination of the time-dependence of the density from the paths is the basic result employed here.

To carry through this programme we need a method of computing the paths for which the Eulerian functions enter only through the initial conditions. Conventionally, computation of the motion of the fluid elements proceeds by specifying an Euler force law for each phase. Then, if the forces depend just on the densities and other known functions, we will obtain a (second-order in time) scheme of the desired type. For our purposes here we proceed in a different way that does not require Euler forces, for we shall assume that the velocity fields are prescribed functions of the $\rho_a$s and other known functions. Eqs. (2.5) then supply a closed system of equations in the $\rho_a$s and the trajectories may be derived from the first order laws (2.4). Conversely, we can solve (2.4) for the trajectories and construct the $\rho_a$s.

## 3. Free wavefunctions

### *3.1 Eulerian picture*

We aim to express the free Schrödinger equation (where $V = 0$ in (1.1)) in the form (2.5) for suitably chosen density and velocity functions. To this end, we replace the complex wavefunction by two real fields: $\psi = \psi_1 + i\psi_2$, $\psi_a \in \Re$, $a = 1,2$. In terms of the real 2-component field $\psi_a(\mathbf{x},t)$ the free wave equation becomes two real equations:

$$\frac{\partial \psi_a}{\partial t} = -\sum_{a'=1}^{2} \frac{\hbar}{2m} \Gamma_{aa'} \nabla^2 \psi_{a'} \tag{3.1}$$

where

$$\Gamma = \begin{pmatrix} 0 & 1 \\ -1 & 0 \end{pmatrix}. \tag{3.2}$$

It is now straightforward to write the wave equation in conservation-equation form:

$$\frac{\partial \psi_a}{\partial t} + \nabla \cdot (\psi_a \mathbf{v}_a) = 0 \tag{3.3}$$

where the velocity fields are

$$\mathbf{v}_a(\mathbf{x},t) = \sum_{a'=1}^{2} \frac{\hbar}{2m} \frac{\Gamma_{aa'} \nabla \psi_{a'}}{\psi_a}, \tag{3.4}$$

i.e.,

$$\mathbf{v}_1(\mathbf{x},t) = \frac{\hbar}{2m} \frac{\nabla \psi_2}{\psi_1}, \quad \mathbf{v}_2(\mathbf{x},t) = -\frac{\hbar}{2m} \frac{\nabla \psi_1}{\psi_2}. \tag{3.5}$$

In fluid language, the free Schrödinger equation may therefore be represented as a two-phase flow whose Eulerian description comprises the brace of real-valued coupled conservation relations (3.3). A peculiarity of the model is that the dynamics of the two fluids is contained entirely in these two equations; the velocity fields of the two phases are determined by the conserved quantities $\rho_1(\mathbf{x},t) = \psi_1(\mathbf{x},t)$ and $\rho_2(\mathbf{x},t) = \psi_2(\mathbf{x},t)$[6]. As anticipated in Sec. 2, no further (force) equations are necessary.

To see how the paths differ from the de Broglie-Bohm ones we rewrite the expressions (3.5) using the polar decomposition $\psi = \sqrt{\rho} \exp(iS/\hbar)$:

$$\mathbf{v}_1 = \frac{\nabla S}{2m} + \frac{\hbar}{4m} \tan(S/\hbar) \nabla \log \rho, \quad \mathbf{v}_2 = \frac{\nabla S}{2m} - \frac{\hbar}{4m} \cot(S/\hbar) \nabla \log \rho. \tag{3.6}$$

---

[6] The interpretation of the 'charge' densities $\psi_a$ is an open problem. They have dimension $(\text{length})^{-3/2}$ and it may prove appropriate to introduce multiplicative constants in these definitions.

This represents the velocities as a special kind of Clebsch representation comprising both the functions $\rho$ and $S$, in contrast to the dynamical law from which the de Broglie-Bohm paths may be derived where these functions enter asymmetrically ($S$ appears only in the initial conditions) [5]. In its first-order form (not the form suitable for computing the wavefunction from the trajectories [5]) the de Broglie-Bohm law is $\mathbf{v} = \nabla S/m$, corresponding to the (single) density $\rho$. Clearly,

$$\cos^2(S/\hbar)\mathbf{v}_1 + \sin^2(S/\hbar)\mathbf{v}_2 = \frac{\nabla S}{2m}. \tag{3.7}$$

The de Broglie-Bohm motion is therefore a kind of local mean over the two phases but the weighting factors do not coincide with the partial densities.

*3.2 Lagrangian picture*

Corresponding to the Eulerian theory of two interacting fields, the Lagrangian viewpoint introduces two sets of spacetime trajectories $q_{1i}(q_{10},t)$ and $q_{2i}(q_{20},t)$, each space point supporting simultaneously a particle of each species. Suppose that at time $t$ the paths $q_{10}$ and $q_{20}$ cross at the point $x$. Then, substituting

$$x_i = q_{1i}(q_{10},t) = q_{2i}(q_{20},t), \quad i=1,2,3, \tag{3.8}$$

into the relations (2.4), (2.7) and (3.5), the coupled Lagrangian flow equations are

$$\frac{\partial q_{1i}(q_{10},t)}{\partial t} = \frac{\hbar}{2m} \frac{1}{J_1^{-1}(q_{10},t)\psi_{10}(q_{10})} J_2^{-1}(q_{20},t) \sum_{j=1}^{3} J_{2ij} \frac{\partial}{\partial q_{20j}} \left[ J_2^{-1}(q_{20},t)\psi_{20}(q_{20}) \right] \tag{3.9}$$

$$\frac{\partial q_{2i}(q_{20},t)}{\partial t} = -\frac{\hbar}{2m} \frac{1}{J_2^{-1}(q_{20},t)\psi_{20}(q_{20})} J_1^{-1}(q_{10},t) \sum_{j=1}^{3} J_{1ij} \frac{\partial}{\partial q_{10j}} \left[ J_1^{-1}(q_{10},t)\psi_{10}(q_{10}) \right] \tag{3.1}$$

where

$$J_{ail} = \sum_{j,k,m,n=1}^{3} \frac{1}{2} \varepsilon_{ijk} \varepsilon_{lmn} \frac{\partial q_{aj}}{\partial q_{a0m}} \frac{\partial q_{ak}}{\partial q_{a0n}}, \quad a=1,2, \quad i,l=1,2,3, \tag{3.11}$$

and $\psi_{10}$ and $\psi_{20}$ are prescribed functions. These relations enable us to compute $q_1(t), q_2(t)$ knowing only the initial wavefunction. Note that whereas the differential equation of motion in the constructive de Broglie-Bohm approach is second order in time and fourth order in particle label [5], doubling the flow reduces the dynamics to first order in time and second order in label.

Conversely, substituting the inverse of (3.8),

$$q_{10i} = q_{10i}(x,t), \quad q_{20i} = q_{20i}(x,t), \tag{3.12}$$

into (3.9) and (3.10), we recover the Eulerian relations (3.5). Employing the Lagrangian viewpoint, therefore, the two sets of paths $q_{ai}(q_{a0},t)$ determine the two real components of the wavefunction at time $t$ according to the formula

$$\psi_a(x,t) = \int \delta(x - q_a(q_{a0},t))\psi_{a0}(q_{a0})d^3q_{a0} \tag{3.13}$$

and the quantum evolution is contained in the statement

$$\psi_a(q_{a0},t)d^3q_a(q_{a0},t) = \psi_{a0}(q_{a0})d^3q_{a0}, a = 1,2. \tag{3.14}$$

The propagator, $K_a(x,t;q_{a0},0) = \delta(x - q_a(q_{a0},t))$, differs from Feynman's, for example, in depending on $\psi_{a0}$.

## 4. Properties of the model

Here we examine some salient features of the two-flow model, referring particularly to further ways in which it differs from the constructive de Broglie-Bohm theory.

(i) *Gauge and Galilean transformations*. The densities $\psi_a$ and velocities (3.5) do not transform as scalars under a global gauge transformation,

$$\begin{pmatrix} \psi_1' \\ \psi_2' \end{pmatrix} = \begin{pmatrix} \cos(ms/\hbar) & -\sin(ms/\hbar) \\ \sin(ms/\hbar) & \cos(ms/\hbar) \end{pmatrix} \begin{pmatrix} \psi_1 \\ \psi_2 \end{pmatrix}, \tag{4.1}$$

i.e., $\rho' = \rho, S' = S + ms$. We shall give examples of distinct trajectory models corresponding to gauge-related wavefunctions in Sec. 5. With a suitable choice of gauge, an instantaneous (non-nodal) singularity in a velocity (3.6) (where $S$ passes through a multiple of $\pi/2$ or $\pi$) may be removed.

In addition, the densities are not scalars, and the velocities are not Galilean 3-vectors, with respect to Galilean boosts:

$$\mathbf{x}' = \mathbf{x} - \mathbf{u}t, \ t' = t, \ \rho' = \rho, \ S' = S + m\mathbf{u}.\mathbf{x} - \tfrac{1}{2}m\mathbf{u}^2 t. \tag{4.2}$$

The non-gauge-, non-Galilean-covariant substratum defined by the sets of variables $q_a(t)$ is nevertheless consistent with the covariance of quantum predictions with respect to these symmetries, as codified in derived quantities such as the density $\rho$. The situation is similar to the non-Lorentz covariant structure of de Broglie-Bohm-like models in the relativistic domain [3] (the non-relativistic de Broglie-Bohm theory is gauge and Galilean covariant). In that case it has been suggested that covariance of the substratum may be restored by including transformations of the particle labels [15]. This possibility remains to be investigated here.

(ii) *Superposition.* For two wavefunctions $\psi, \bar{\psi}$, the velocities corresponding to the superposition $\tilde{\psi} = \psi + \bar{\psi}$ are

$$\tilde{\mathbf{v}}_1 = \frac{\psi_1 \mathbf{v}_1 + \bar{\psi}_1 \bar{\mathbf{v}}_1}{\psi_1 + \bar{\psi}_1}, \quad \tilde{\mathbf{v}}_2 = \frac{\psi_2 \mathbf{v}_2 + \bar{\psi}_2 \bar{\mathbf{v}}_2}{\psi_2 + \bar{\psi}_2}. \tag{4.3}$$

In contrast to the de Broglie-Bohm velocity [3], these expressions contain no interference terms; they are simple averages with the corresponding fractional densities as weights.

(iii) *One-phase flow.* The trajectories of the two phases will coincide when
$\mathbf{v}_1 = \mathbf{v}_2$ for all $x,t$
which implies $\nabla \rho = 0$ or $\rho = \rho(t)$. Examples are a plane wave and the free Green function.

(iv) *Many-body systems.* The model extends in an obvious way to a system of particles by increasing the range of the indices; there are two velocity fields in the configuration space and the wavefunction is constructed according to the formula (3.14) with $d^3 q_{a0}$ replaced by $d^n q_{a0}$ etc. A noteworthy feature of this theory is that the Lagrangian trajectories of each body depend irreducibly on the total quantum state. Thus, for a separable state the paths corresponding to one body do not depend just on the quantum state associated with that body.

To see this, consider two bodies of equal mass where the state is a product:

$$\varphi(\mathbf{x}_1, \mathbf{x}_2) = \psi(\mathbf{x}_1)\phi(\mathbf{x}_2). \tag{4.4}$$

Writing $\psi = \psi_1 + i\psi_2, \phi = \phi_1 + i\phi_2$, the two real fields associated with the total system are entangled functions of the two sets of real fields corresponding to each component system:

$$\varphi = \varphi_1 + i\varphi_2 = (\psi_1 \phi_1 - \psi_2 \phi_2) + i(\psi_1 \phi_2 + \psi_2 \phi_1). \tag{4.5}$$

Then the velocity 6-vectors are

$$\mathbf{v}_1(\mathbf{x}_1, \mathbf{x}_2) = \frac{\hbar}{2m} \frac{\nabla \varphi_2}{\varphi_1}, \quad \mathbf{v}_2(\mathbf{x}_1, \mathbf{x}_2) = -\frac{\hbar}{2m} \frac{\nabla \varphi_1}{\varphi_2} \tag{4.6}$$

where $\nabla$ is a six-dimensional gradient and we have in general

$$v_{1i}(\mathbf{x}_1,\mathbf{x}_2) \neq \frac{\hbar}{2m}\frac{\partial_i \psi_2}{\psi_1}, \quad v_{2i}(\mathbf{x}_1,\mathbf{x}_2) \neq -\frac{\hbar}{2m}\frac{\partial_i \psi_1}{\psi_2}, \quad i=1,2,3. \tag{4.7}$$

Unlike the analogous expression for velocity in the de Broglie-Bohm theory where separable states imply independent motions, the coordinates of particle 2 do not drop out of the problem and there is, in general, no simple relation between the velocity computed from the total wavefunction ($\varphi$) and that from the single ($\psi$). This feature presents no conceptual problem for the construction and we may work with either the total system or the individual ones to obtain $\varphi$. Naturally, the 'entanglement' of the paths in separable states is consistent with the statistical independence expressed by (4.4).

## 5. External field case: conservation in five dimensions

### *5.1 Eulerian picture*

The external scalar potential $V(\mathbf{x},t)$ in the Schrödinger equation (1.1) adds a field-dependent source term to the continuity equation (3.3). In order to pursue our programme of writing field equations in conservation-like form without sources, we bring in the external term as part of a free equation in a higher-dimensional Riemannian space, *à la* Kaluza-Klein. To this end, we employ the metric used in Eisenhart's geometrization of classical mechanics in which Newton's force law is represented as a geodesic in a five-dimensional curved space (the Riemann tensor being connected with the stability of trajectories) [16-19]. In the analogous problem for quantum mechanics, the same technique may be used to express the Schrödinger equation (1.1) as a free wave equation in the five-dimensional Eisenhart geometry [20, 21]. As a by-product, this procedure puts the quantum dynamics in the desired conservation form (2.5). This construction is interesting also in giving alternative models to that of Sec. 3 in the special case $V=0$.

Denoting the coordinates on the five-dimensional space by $x^\alpha = (t,\mathbf{x},s)$, $\alpha = 0,1,2,3,4, -\infty \leq s \leq \infty$, the massless wave equation is

$$\sum_{\alpha,\beta=0}^{4} \left(1/\sqrt{-g}\right)\partial_\alpha\left(\sqrt{-g}\, g^{\alpha\beta}\partial_\beta \phi\right) = 0 \tag{5.1}$$

where $g = \det g_{\alpha\beta}$. Inserting for $g_{\alpha\beta}$ the Eisenhart metric defined by

$$\sum_{\alpha,\beta=0}^{4} g_{\alpha\beta}dx^\alpha dx^\beta = \sum_{\mu,\nu=1}^{3} \delta_{\mu\nu}dx^\mu dx^\nu + dsdt + dtds - \left[2V(\mathbf{x},t)/m\right]dt^2, \tag{5.2}$$

for which $g = -1$, we obtain

$$\frac{\partial^2 \phi}{\partial t \partial s} + \frac{1}{2}\nabla^2 \phi + \frac{V}{m}\frac{\partial^2 \phi}{\partial s^2} = 0. \tag{5.3}$$

Assuming that $\phi$ is an eigenstate of the 5th component of the momentum,

$$-i\hbar \frac{\partial \phi}{\partial s} = m\phi \text{ or } \phi(\mathbf{x},t,s) = e^{ims/\hbar}\psi(\mathbf{x},t), \tag{5.4}$$

yields Schrödinger's equation (1.1). Conversely, starting with (1.1), the function $\phi$ defined in (5.4) obeys the five-dimensional equation (5.1) with metric (5.2).

Using the real components $\phi_a$, $a = 1,2$, (5.3) is equivalent to two identical real equations:

$$\frac{\partial^2 \phi_a}{\partial t \partial s} + \frac{1}{2}\nabla^2 \phi_a + \frac{V}{m}\frac{\partial^2 \phi_a}{\partial s^2} = 0. \tag{5.5}$$

When $V = 0$ this gives an alternative formulation of the free Schrödinger equation to that of Sec. 3 in that each component $\phi_a$ evolves independently. It is straightforward to write (5.5) as a continuity equation for each $a$:

$$\frac{\partial}{\partial t}\frac{\partial \phi_a}{\partial s} + \nabla \cdot \left(\frac{\partial \phi_a}{\partial s}\bar{\mathbf{v}}_a\right) + \frac{\partial}{\partial s}\left(\frac{\partial \phi_a}{\partial s}u_a\right) = 0 \tag{5.6}$$

where the velocity fields are

$$\bar{\mathbf{v}}_a(\mathbf{x},t,s) = \frac{\nabla \phi_a}{2\partial \phi_a/\partial s}, \quad u_a(\mathbf{x},t,s) = \frac{V}{m}. \tag{5.7}$$

The conserved quantities are therefore $\partial \phi_a/\partial s$. However, the five-dimensional theory is not satisfactory in this form since the velocities are not functions of just the conserved quantities and the prescribed function $V$ ($\bar{\mathbf{v}}_a$ involves $\nabla \phi_a$). To obtain the desired formulation we must relinquish the independent evolution of the real components of the wavefunction. Using the constraint (5.4) we shall instead write (5.5) as

$$i\hbar \frac{\partial \phi}{\partial t} = -\frac{\hbar^2}{2m}\nabla^2 \phi - \frac{\hbar^2}{m^2}V\frac{\partial^2 \phi}{\partial s^2}. \tag{5.8}$$

This is equivalent to the two real equations

$$\frac{\partial \phi_a}{\partial t} = -\sum_{a'=1}^{2}\frac{\hbar}{2m}\Gamma_{aa'}\nabla^2 \phi_{a'} - \sum_{a'=1}^{2}\frac{\hbar}{m^2}V\Gamma_{aa'}\frac{\partial^2 \phi_{a'}}{\partial s^2}. \tag{5.9}$$

In this version the conservation-equation form is

$$\frac{\partial \phi_a}{\partial t} + \nabla \cdot (\phi_a \mathbf{v}_a) + \frac{\partial}{\partial s}(\phi_a u_a) = 0 \tag{5.10}$$

where the velocity fields are

$$\mathbf{v}_a(\mathbf{x},s,t) = \sum_{a'=1}^{2} \frac{\hbar}{2m} \frac{\Gamma_{aa'} \nabla \phi_{a'}}{\phi_a}, \quad u_a(\mathbf{x},s,t) = \sum_{a'=1}^{2} \frac{\hbar}{m^2} V \frac{\Gamma_{aa'} \partial \phi_{a'}/\partial s}{\phi_a} \tag{5.11}$$

or

$$\begin{aligned}
\mathbf{v}_1 &= \frac{\hbar}{2m} \frac{\nabla \phi_2}{\phi_1}, & u_1 &= \frac{\hbar}{m^2} V \frac{\partial \phi_2/\partial s}{\phi_1}, \\
\mathbf{v}_2 &= -\frac{\hbar}{2m} \frac{\nabla \phi_1}{\phi_2}, & u_2 &= -\frac{\hbar}{m^2} V \frac{\partial \phi_1/\partial s}{\phi_2}.
\end{aligned} \tag{5.12}$$

Now the velocities are functions of the conserved quantities $\phi_a$, and $V$, alone.

*5.2 Lagrangian picture*

The 4+1-dimensional theory follows closely the Euclidean space free theory of Sec. 3.2. We introduce two sets of trajectories $q_{1i}(q_{10},t)$ and $q_{2i}(q_{20},t)$ where $q_{10}(q_{20})$ represents the initial values of $q_{1i}(q_{2i})$, $i=1,2,3,4$. If at time $t$ the paths $q_{10}$ and $q_{20}$ cross at the point $(\mathbf{x},s)$, the Lagrangian equations of motion obtained from (5.12) are

$$\frac{\partial q_{1i}(q_{10},t)}{\partial t} = \frac{\hbar}{2m} \frac{1}{J_1^{-1}(q_{10},t)\phi_{10}(q_{10})} J_2^{-1}(q_{20},t) \sum_{j=1}^{4} J_{2ij} \frac{\partial}{\partial q_{20j}} \left[ J_2^{-1}(q_{20},t)\phi_{20}(q_{20}) \right] \tag{5.13}$$

$$\frac{\partial q_{2i}(q_{20},t)}{\partial t} = -\frac{\hbar}{2m} \frac{1}{J_2^{-1}(q_{20},t)\phi_{20}(q_{20})} J_1^{-1}(q_{10},t) \sum_{j=1}^{4} J_{1ij} \frac{\partial}{\partial q_{10j}} \left[ J_1^{-1}(q_{10},t)\phi_{10}(q_{10}) \right] \tag{5.14}$$

where

$$J_{aim} = \sum_{j,k,l,n,p,r=1}^{4} \frac{1}{6} \varepsilon_{ijkl} \varepsilon_{mnpr} \frac{\partial q_{aj}}{\partial q_{a0n}} \frac{\partial q_{ak}}{\partial q_{a0p}} \frac{\partial q_{al}}{\partial q_{a0r}}, \quad a=1,2, \quad i,m=1,2,3,4. \tag{5.15}$$

The Lagrangian equivalent of the constraint (5.4) fixes the initial wavefunction:

$$\begin{aligned}
\phi_{10}(q_{10}) &= \cos(mq_{104}/\hbar)\psi_{10}(q_{10i}) - \sin(mq_{104}/\hbar)\psi_{20}(q_{10i}), \\
\phi_{20}(q_{20}) &= \sin(mq_{204}/\hbar)\psi_{10}(q_{20i}) + \cos(mq_{204}/\hbar)\psi_{20}(q_{20i}), \quad i=1,2,3.
\end{aligned} \tag{5.16}$$

It also simplifies the equations for $q_{a4}$:

$$\frac{\partial q_{a4}}{\partial t} = \frac{1}{m} V(q_{ai}(q_{a0i},t),t) \implies q_{a4} = \frac{1}{m}\int_0^t V(q_{ai}(q_{a0i},t),t)dt + q_{a04}, \quad i=1,2,3, \quad (5.17)$$

which may be substituted in the remaining equations (5.13) and (5.14).

The evolution of a quantum system in an external potential is then encoded in the conservation law

$$\phi_a(q_{a0},t)d^4q_a(q_{a0},t) = \phi_{a0}(q_{a0})d^4q_{a0}, \quad a=1,2, \quad (5.18)$$

and in propagator form it is

$$\phi_a(x_i,s,t) = \int \delta(x_i - q_{ai}(q_{a0},t))\delta(s - q_{a4}(q_{a0},t))\phi_{a0}(q_{a0})d^4q_{a0}, \quad i=1,2,3. \quad (5.19)$$

The function $\psi$ may be extracted by inverting (5.4).

When $V = 0$, we have $q_{a4}(q_{a0},t) = q_{a04}$, $a=1,2$, and so (5.18) reduces to a 3+1-dimensional relation of the type (3.14). The theory does not coincide with that of Sec. 3, however, since the wavefunction $\phi$ is a gauge transform of $\psi$ used there.

**6. Alternative two-phase models**

There are other ways of writing the Schrödinger equation in conservation form, in addition to those presented above, which imply alternative two-phase flows and corresponding methods of state construction. We shall illustrate three possibilities, just for the free case, distinguished by the choice of density functions: in the first case the densities are as used above, in the second case one density is different, and in the third case both are different. The second example features the de Broglie-Bohm trajectories in a novel setting.

To begin with, we may retain the densities $\psi_a$, $a=1,2$, but associate them with different flows, if we define new velocities by adding to each current $\psi_a v_a$ a divergence-free term that depends just on the density. An example of such an addition, inspired by a spin-dependent term required in the de Broglie-Bohm theory [7], is $\nabla \psi_a \times \mathbf{w}_a$ where $\mathbf{w}_a$ is a constant vector. Then

$$\mathbf{v}_a = \sum_{a'=1}^{2}\frac{\hbar}{2m}\frac{\Gamma_{aa'}\nabla\psi_{a'}}{\psi_a} + \frac{\nabla\psi_a \times \mathbf{w}_a}{\psi_a}, \quad a=1,2. \quad (6.1)$$

The second example is provided by the observation that the free Schrödinger equation is equivalent to one of eqs. (3.3), say $a = 1$, together with the usual conservation equation

$$\frac{\partial \rho}{\partial t} + \nabla \cdot (\rho \mathbf{v}) = 0. \tag{6.2}$$

This implies a two-phase model in which the densities are identified as $\rho_1 = \psi_1, \rho_2 = \psi_1^2 + \psi_2^2$. Then, once again, the velocities may be expressed purely in terms of the densities but this time the association is

$$\mathbf{v}_1 = \frac{\hbar}{2m} \frac{\nabla \pm (\rho_2 - \rho_1^2)^{1/2}}{\rho_1}, \mathbf{v}_2 = \frac{\hbar}{m} \nabla \tan^{-1} \frac{\pm (\rho_2 - \rho_1^2)^{1/2}}{\rho_1}. \tag{6.3}$$

The velocity $\mathbf{v}_2$ is just the de Broglie-Bohm velocity. The Lagrangian version of these relations implies an alternative method of computing the de Broglie-Bohm paths to the second-order method described previously [5]. The wavefunction may be found from the formulas $\psi_1 = \rho_1, \psi_2 = \pm(\rho_2 - \rho_1^2)^{1/2}$.

The third example follows from the observation that the linear combinations $\rho_1 = \alpha \psi_1 + \beta \psi_2, \rho_2 = \alpha \psi_1 - \beta \psi_2$ ($\alpha, \beta$ = real constants ) of the original densities obey conservation equations corresponding to the velocity fields

$$\mathbf{v}_1 = \frac{\hbar}{4m\rho_1}(\gamma \nabla \rho_1 - \delta \nabla \rho_2), \quad \mathbf{v}_2 = \frac{\hbar}{4m\rho_2}(\delta \nabla \rho_1 - \gamma \nabla \rho_2), \tag{6.4}$$

where $\gamma = (\alpha^2 - \beta^2)/\alpha\beta, \delta = (\alpha^2 + \beta^2)/\alpha\beta$. The wavefunction in this case is given by $\psi_1 = (\rho_1 + \rho_2)/2\alpha, \psi_2 = (\rho_1 - \rho_2)/2\beta$.

## 7. A cornucopia of trajectory theories

In the last 15 years or so it has been established that there are several ways of introducing the trajectory concept into quantum mechanics. The purpose and indeed the value of the various proposals varies. Trajectories have been invoked with the aim of (i) interpreting quantum mechanics (as 'hidden variables', e.g., de Broglie-Bohm), (ii) constructing or computing the wavefunction (e.g., the method described here), and (iii) providing insight into specific aspects of the formalism (e.g., the energy flow lines of the Schrödinger field defined by the energy-momentum complex [3]). These categories may overlap.

A common theme of many of these proposals is that the paths are connected to the wavefunction via conservation-like equations. It has been argued (in the context of recent work on complex trajectories) that for computational purposes it is essential that the ensemble of trajectories obeys the quantal probability conservation equation [22]. However, whilst it may be computationally efficient to have the trajectories follow the probability, this seems to be an issue of pragmatism rather than of principle. It all depends on the purposes to which the trajectories are to be put. The key point in generating $\psi(t)$ from $\psi_0$ via sets of paths in a fluid-mechanical context is that one has *a* conservation equation, or equations, supplying sufficient propagation information; $\rho d^3 q$ need not be an invariant of the flow. In the theory of this paper, for example, the conserved densities are the functions $\psi_a$ rather than the single density $\rho$. There are other examples, such as the trajectories that propagate the Wigner function (which of course can take negative as well as positive values) [6], or those connected with unconventional probability distributions [23].

The non-uniqueness of quantum trajectory representations need not be regarded as a defect to be remedied by a 'proof' of the 'true' picture. Rather, it indicates that a field theory (wave mechanics) can be represented, in fluid terms, through a variety of non-trivially distinct Eulerian models, depending on the identification of the density, velocity and other functions. To each of the Eulerian formulations there may correspond a distinct Lagrangian model. These models may be regarded as 'equivalent', insofar as the various sets of fluid functions in each map into Schrödinger dynamics, yet they provide a range of mathematical tools and insights into quantum processes. The possibility of developing a comparable range of formulations of the laws of classical continuum mechanics does not seem to have been widely appreciated.

There is further work to do in elucidating possible connections between hydrodynamic-type theories and other constructive theories that employ spacetime trajectories as building blocks of quantum propagation, such as the path integral formalism. It may be propitious to regard the two-path model of this article as occupying an intermediate position in a spectrum of theories bounded by Feynman (points are linked by all possible paths) and de Broglie-Bohm and related models (points are linked by at most one path), which prompts an enquiry into whether there are other finite-path models, and whether the path integral might be treated as a multiphase flow.